\documentclass[%
 reprint,
nofootinbib,
 amsmath,amssymb,
 aps,
]{revtex4-2}
\usepackage{graphicx}
\usepackage{bm}
\usepackage{physics}
\usepackage{lineno}

\newcommand{\qcl}{\mathbf{q}\times\mathbf{l}}
\newcommand{\Ph}{\mathbf{P}_h}
\newcommand{\Rperp}{\mathbf{R}_{\perp}}

\newcommand{\qcPh}{\mathbf{q}\times\Ph}
\newcommand{\qcRperp}{\mathbf{q}\times\Rperp}

\begin{document}
\preprint{APS/123-QED}

\title{Measurements of Beam Spin Asymmetries of $\pi^\pm\pi^0$ dihadrons at CLAS12}

\newcommand*{\ANL}{Argonne National Laboratory, Argonne, Illinois 60439}
\newcommand*{\ANLindex}{1}
\affiliation{\ANL}
\newcommand*{\SACLAY}{IRFU, CEA, Universit'{e} Paris-Saclay, F-91191 Gif-sur-Yvette, France}
\newcommand*{\SACLAYindex}{2}
\affiliation{\SACLAY}
\newcommand*{\CNU}{Christopher Newport University, Newport News, Virginia 23606}
\newcommand*{\CNUindex}{3}
\affiliation{\CNU}
\newcommand*{\UCONN}{University of Connecticut, Storrs, Connecticut 06269}
\newcommand*{\UCONNindex}{4}
\affiliation{\UCONN}
\newcommand*{\DUKE}{Duke University, Durham, North Carolina 27708-0305}
\newcommand*{\DUKEindex}{5}
\affiliation{\DUKE}
\newcommand*{\DUQUESNE}{Duquesne University, 600 Forbes Avenue, Pittsburgh, PA 15282 }
\newcommand*{\DUQUESNEindex}{6}
\affiliation{\DUQUESNE}
\newcommand*{\FU}{Fairfield University, Fairfield CT 06824}
\newcommand*{\FUindex}{7}
\affiliation{\FU}
\newcommand*{\FERRARAU}{Universita' di Ferrara , 44121 Ferrara, Italy}
\newcommand*{\FERRARAUindex}{8}
\affiliation{\FERRARAU}
\newcommand*{\FIU}{Florida International University, Miami, Florida 33199}
\newcommand*{\FIUindex}{9}
\affiliation{\FIU}
\newcommand*{\GWUI}{The George Washington University, Washington, DC 20052}
\newcommand*{\GWUIindex}{10}
\affiliation{\GWUI}
\newcommand*{\GSIFFN}{GSI Helmholtzzentrum fur Schwerionenforschung GmbH, D-64291 Darmstadt, Germany}
\newcommand*{\GSIFFNindex}{11}
\affiliation{\GSIFFN}
\newcommand*{\ORSAY}{Universit'{e} Paris-Saclay, CNRS/IN2P3, IJCLab, 91405 Orsay, France}
\newcommand*{\ORSAYindex}{12}
\affiliation{\ORSAY}
\newcommand*{\INFNFE}{INFN, Sezione di Ferrara, 44100 Ferrara, Italy}
\newcommand*{\INFNFEindex}{13}
\affiliation{\INFNFE}
\newcommand*{\INFNFR}{INFN, Laboratori Nazionali di Frascati, 00044 Frascati, Italy}
\newcommand*{\INFNFRindex}{14}
\affiliation{\INFNFR}
\newcommand*{\INFNGE}{INFN, Sezione di Genova, 16146 Genova, Italy}
\newcommand*{\INFNGEindex}{15}
\affiliation{\INFNGE}
\newcommand*{\INFNRO}{INFN, Sezione di Roma Tor Vergata, 00133 Rome, Italy}
\newcommand*{\INFNROindex}{16}
\affiliation{\INFNRO}
\newcommand*{\INFNTUR}{INFN, Sezione di Torino, 10125 Torino, Italy}
\newcommand*{\INFNTURindex}{17}
\affiliation{\INFNTUR}
\newcommand*{\INFNPAV}{INFN, Sezione di Pavia, 27100 Pavia, Italy}
\newcommand*{\INFNPAVindex}{18}
\affiliation{\INFNPAV}
\newcommand*{\KNU}{Kyungpook National University, Daegu 41566, Republic of Korea}
\newcommand*{\KNUindex}{19}
\affiliation{\KNU}
\newcommand*{\LAMAR}{Lamar University, 4400 MLK Blvd, PO Box 10046, Beaumont, Texas 77710}
\newcommand*{\LAMARindex}{20}
\affiliation{\LAMAR}
\newcommand*{\MIT}{Massachusetts Institute of Technology, Cambridge, Massachusetts  02139-4307}
\newcommand*{\MITindex}{21}
\affiliation{\MIT}
\newcommand*{\MISS}{Mississippi State University, Mississippi State, MS 39762-5167}
\newcommand*{\MISSindex}{22}
\affiliation{\MISS}
\newcommand*{\UNH}{University of New Hampshire, Durham, New Hampshire 03824-3568}
\newcommand*{\UNHindex}{23}
\affiliation{\UNH}
\newcommand*{\NMSU}{New Mexico State University, PO Box 30001, Las Cruces, NM 88003, USA}
\newcommand*{\NMSUindex}{24}
\affiliation{\NMSU}
\newcommand*{\OHIOU}{Ohio University, Athens, Ohio  45701}
\newcommand*{\OHIOUindex}{25}
\affiliation{\OHIOU}
\newcommand*{\ODU}{Old Dominion University, Norfolk, Virginia 23529}
\newcommand*{\ODUindex}{26}
\affiliation{\ODU}
\newcommand*{\JLUGiessen}{II Physikalisches Institut der Universitaet Giessen, 35392 Giessen, Germany}
\newcommand*{\JLUGiessenindex}{27}
\affiliation{\JLUGiessen}
\newcommand*{\URICH}{University of Richmond, Richmond, Virginia 23173}
\newcommand*{\URICHindex}{28}
\affiliation{\URICH}
\newcommand*{\ROMAII}{Universita' di Roma Tor Vergata, 00133 Rome Italy}
\newcommand*{\ROMAIIindex}{29}
\affiliation{\ROMAII}
\newcommand*{\SDU}{Shandong University, Qingdao, Shandong 266237, China}
\newcommand*{\SDUindex}{30}
\affiliation{\SDU}
\newcommand*{\MSU}{Skobeltsyn Institute of Nuclear Physics, Lomonosov Moscow State University, 119234 Moscow, Russia}
\newcommand*{\MSUindex}{31}
\affiliation{\MSU}
\newcommand*{\SCAROLINA}{University of South Carolina, Columbia, South Carolina 29208}
\newcommand*{\SCAROLINAindex}{32}
\affiliation{\SCAROLINA}
\newcommand*{\TEMPLE}{Temple University,  Philadelphia, PA 19122 }
\newcommand*{\TEMPLEindex}{33}
\affiliation{\TEMPLE}
\newcommand*{\JLAB}{Thomas Jefferson National Accelerator Facility, Newport News, Virginia 23606}
\newcommand*{\JLABindex}{34}
\affiliation{\JLAB}
\newcommand*{\ULS}{Universidad de La Serena}
\newcommand*{\ULSindex}{35}
\affiliation{\ULS}
\newcommand*{\UTFSM}{Universidad T\'{e}cnica Federico Santa Mar\'{i}a, Casilla 110-V Valpara\'{i}so, Chile}
\newcommand*{\UTFSMindex}{36}
\affiliation{\UTFSM}
\newcommand*{\INSUBRIA}{Universit\`{a} degli Studi dell'Insubria, 22100 Como, Italy}
\newcommand*{\INSUBRIAindex}{37}
\affiliation{\INSUBRIA}
\newcommand*{\BRESCIA}{Universit`{a} degli Studi di Brescia, 25123 Brescia, Italy}
\newcommand*{\BRESCIAindex}{38}
\affiliation{\BRESCIA}
\newcommand*{\GLASGOW}{University of Glasgow, Glasgow G12 8QQ, United Kingdom}
\newcommand*{\GLASGOWindex}{39}
\affiliation{\GLASGOW}
\newcommand*{\YORK}{University of York, York YO10 5DD, United Kingdom}
\newcommand*{\YORKindex}{40}
\affiliation{\YORK}
\newcommand*{\VIRGINIA}{University of Virginia, Charlottesville, Virginia 22901}
\newcommand*{\VIRGINIAindex}{41}
\affiliation{\VIRGINIA}
\newcommand*{\WM}{College of William and Mary, Williamsburg, Virginia 23187-8795}
\newcommand*{\WMindex}{42}
\affiliation{\WM}
\newcommand*{\YEREVAN}{Yerevan Physics Institute, 375036 Yerevan, Armenia}
\newcommand*{\YEREVANindex}{43}
\affiliation{\YEREVAN}
\newcommand*{\JMU}{James Madison University, Harrisonburg, Virginia 22807}
\newcommand*{\JMUindex}{44}
\affiliation{\JMU}

\newcommand*{\NOWJLAB}{Thomas Jefferson National Accelerator Facility, Newport News, Virginia 23606}

\author {A.G.~Acar} 
\affiliation{\YORK}
\author {P.~Achenbach} 
\affiliation{\CNU}
\author {J. S. Alvarado} 
\affiliation{\ORSAY}
\author {M.~Amaryan} 
\affiliation{\ODU}
\author {W.R.~Armstrong} 
\affiliation{\ANL}
\author {H.~Atac} 
\affiliation{\TEMPLE}
\author {H.~Avakian} 
\affiliation{\JLAB}
\author {N.A.~Baltzell} 
\affiliation{\JLAB}
\author {L. Barion} 
\affiliation{\INFNFE}
\author {M.~Battaglieri} 
\affiliation{\INFNGE}
% ...

\collaboration{The CLAS Collaboration}
\noaffiliation
\date{\today}

\begin{abstract}
A first measurement of beam spin asymmetries for $\pi^+\pi^0$ and $\pi^-\pi^0$ pairs in semi-inclusive deep inelastic scattering is reported. The asymmetries in the dihadron angular distributions were measured from the scattering of a 10.6 GeV longitudinally polarized electron beam off a proton target, using the CLAS12 detector at Jefferson Lab. A photon classifier using a Gradient Boosted Trees (GBTs) architecture was trained with Monte Carlo simulations to reduce the amount of false combinatorial background $\pi^0$s, increasing statistics by up to five-fold compared to previous CLAS12 $\pi^0$ analyses. A nonzero $\sin\phi_{R_\perp}$ asymmetry is observed. This measurement is sensitive to the underexplored collinear twist-3 PDF $e(x)$, which encodes quark-gluon correlations in the proton, and presents a new avenue for its point-by-point extraction. The asymmetries also provide the first experimental evidence for the isospin-dependence of the helicity-dependent dihadron fragmentation function $G_1^\perp$, revealed by a sign-difference between the $\pi^+\pi^0$ and $\pi^-\pi^0$ channels in the $\sin(\phi_h-\phi_{R_\perp})$ modulation. In contrast, a large, same-sign enhancement near the $\rho$ mass for the $\sin(2\phi_h-2\phi_{R_\perp})$ modulation is observed, matching spectator model predictions in $\pi^+\pi^-$ pairs. 

\end{abstract}

\maketitle
Nucleons and electrons form the building blocks of most visible matter in our Universe. Whereas electrons are considered point-like in nature, nucleons have a more complex internal structure consisting of quarks and gluons (collectively partons) bound through gluonic interactions described by the strong force. It is still unknown how these interactions create the emergent properties of a proton, such as its mass and spin~\cite{accardi2014electronioncolliderqcd}. By using the proton as our laboratory for studying partonic interactions, we will improve our understanding of the strong force and the matter around us.

Parton Distribution Functions (PDFs) encode elements of nucleon structure and are measured in scattering experiments across a range of kinematic conditions. PDFs, along with other Quantum Chromodynamics (QCD) objects, can be categorized by their ``twist". In QCD, the twist of a distribution is defined as its mass dimension minus its spin, with higher twist terms representing power-suppressed multi-parton correlations. Leading twist (twist-2) PDFs like $f_1(x)$ give the probability of finding a parton with momentum fraction $x$ in the nucleon~\cite{Collins_2011}. However, a key limitation of twist-2 PDFs is that they reveal no information about the quark-gluon correlations thought to generate the emergent properties of nucleons.

To probe these internal dynamics experimentally, one requires higher twist PDFs~\cite{Jaffe_Ji_1992}. Twist-3 PDFs capture quark-gluon interactions, though they come with a $1/Q$ suppression in the scattering cross section relative to leading twist. Among the twist-3 collinear PDFs $g_T(x)$, $h_L(x)$ and $e(x)$, the least known is $e(x)$, having only been recently extracted for the first time~\cite{Courtoy_2022} using data collected at the \mbox{CEBAF} Large Acceptance Spectrometer (CLAS)~\cite{PhysRevLett.126.062002} and its 12 GeV upgrade CLAS12~\cite{PhysRevLett.126.152501}. Unlike $g_T(x)$ and $h_L(x)$ which have twist-2 counterparts, $e(x)$ does not~\cite{Wandzura_Wilczek_1977}. As a result, Wandzura–Wilczek-type approximations cannot be applied, making direct experimental measurement one of the few viable methods to access it. 

Beyond this, $e(x)$ is of particular scientific interest as its $x$-moments are related to fundamental properties of nucleons. Its first moment is proportional to the quark contribution to the nucleon $\sigma$-term, linking $e(x)$ directly to mechanisms of dynamical chiral symmetry breaking and hadron mass generation~\cite{Cebulla_Ossmann_Schweitzer_2007}. The flavor-singlet combination $\int dx\,\left(e^u(x) + e^d(x)\right)$, where $u$ and $d$ represent up and down quarks, is related to the \mbox{pion-nucleon} \mbox{$\sigma$-term}, a quantity central to chiral perturbation theory and of phenomenological relevance in models of dark matter-nucleon scattering~\cite{PhysRevD.77.065026}. Additionally, the second moment of $e(x)$ encodes information about the transverse force acting on a polarized quark as it exits the nucleon, providing insight into nontrivial aspects of QCD dynamics~\cite{burkardt2013transverseforcequarksdis}.

By comparison, the non-perturbative dynamics of hadronization, the QCD process where quarks and gluons combine to form bound state hadrons, is described using fragmentation functions (FFs). Similar to PDFs, leading twist FFs have a probabilistic interpretation, describing the likelihood of quark $q$ to form a hadron $h$. 

Semi-inclusive deep inelastic scattering (SIDIS) is a popular tool to access PDFs and FFs. However, extracting $e(x)$ from single hadron SIDIS is challenging because it appears in the beam spin asymmetry (BSA) as a linear combination of four convoluted PDF$\otimes$FF pairs, and each PDF and FF must be parameterized with model-based transverse momentum-dependence, introducing theoretical uncertainties.

Both issues are resolved by extracting $e(x)$ using dihadron SIDIS, where two hadrons are measured in the final state. In this study, we consider the production of $\pi^+\pi^0$ and $\pi^-\pi^0$ dihadrons from electron-proton scattering:
\begin{equation}\label{eq:dihadronSidisEqn}
    e^-(\ell)+p(P)\rightarrow e^-(\ell')+h_1(P_1)+h_2(P_2)+X,
\end{equation}
where the first hadron $h_1$ refers to the charged pion $\pi^\pm$, the second hadron $h_2$ refers to the neutral pion $\pi^0$, and the quantities in the parentheses are four-momenta. Since the $\pi^0$ decays into two photons electromagnetically, we measure it experimentally by reconstructing two photon (diphoton) pairs (i.e. $P_2=P_{\gamma1}+P_{\gamma2}$).

The dihadron SIDIS framework models hadronization with dihadron fragmentation functions (DiFFs), which at leading twist give the probability for a quark $q$ to fragment into a hadron pair $h_1h_2$. They depend on the fraction of the fragmenting quark momentum carried by the pion pair $z$, the invariant mass of the pion pair $M_h$, and the angle $\theta$ between $\mathbf{P}_1$ in the dihadron center-of-mass frame and $\mathbf{P}_h\equiv\mathbf{P}_{1}+\mathbf{P}_2$ in the photon-target rest frame (which is integrated over in this work). A partial wave expansion of DiFFs in terms of $\theta$~\cite{Gliske_Bacchetta_Radici_2014,bacchetta_two-hadron_2004} can be performed using associated Legendre polynomials. The additional degrees of freedom allow the hadron pair to carry measurable relative transverse momentum. The chiral-odd DiFF $H_1^\sphericalangle$ can couple to $e(x)$ in the dihadron SIDIS beam spin asymmetry without explicit quark transverse-momentum dependence, allowing for point-by-point extraction of $e(x)$ without modeling~\cite{bacchetta_two-hadron_2004,Courtoy_2022}.

Furthermore, dihadron SIDIS allows one to study new complex hadronization correlations that are not present in single hadron SIDIS. The helicity-dependent DiFF $G_1^\perp$ represents a correlation between a fragmenting quark's longitudinal polarization and the angular distribution of the produced hadron pair. There is no analog to $G_1^\perp$ in single hadron SIDIS, and it has only been measured to be nonzero for the first time at CLAS12~\cite{PhysRevLett.126.152501}. Recent developments modeling the behavior of $G_1^\perp$~\cite{Luo_Sun_2020,Luo_Sun_Xie_2020} for fragmented $\pi^+\pi^-$ pairs predict enhancements near the $\rho^0$ mass and have been validated experimentally.


Measurements of the $\pi^\pm\pi^0$ dihadron channel were carried out at Jefferson Lab using the CLAS12 detector
\cite{Burkert_Elouadrhiri_Adhikari_Adhikari_Amaryan_Anderson_Angelini_Antonioli_Atac_Aune_etal._2020}. Between Spring 2018 and Spring 2019, the experiment scattered a longitudinally polarized electron beam with an energy of 10.2-10.6 GeV and run-by-run polarization of 86–89\% on a fixed, unpolarized liquid hydrogen target. The analysis measured electrons, pions, and photons using CLAS12's Forward Detector, a subsystem of Cherenkov counters~\cite{UNGARO2020163420,SHARABIAN2020163824}, drift chamber tracking detectors~\cite{MESTAYER2020163518}, and electromagnetic calorimeters (ECals)~\cite{ASRYAN2020163425} spanning a scattering angle range of $5-35^\circ$. 

\begin{figure}[ht]
\includegraphics[width=0.48\textwidth]{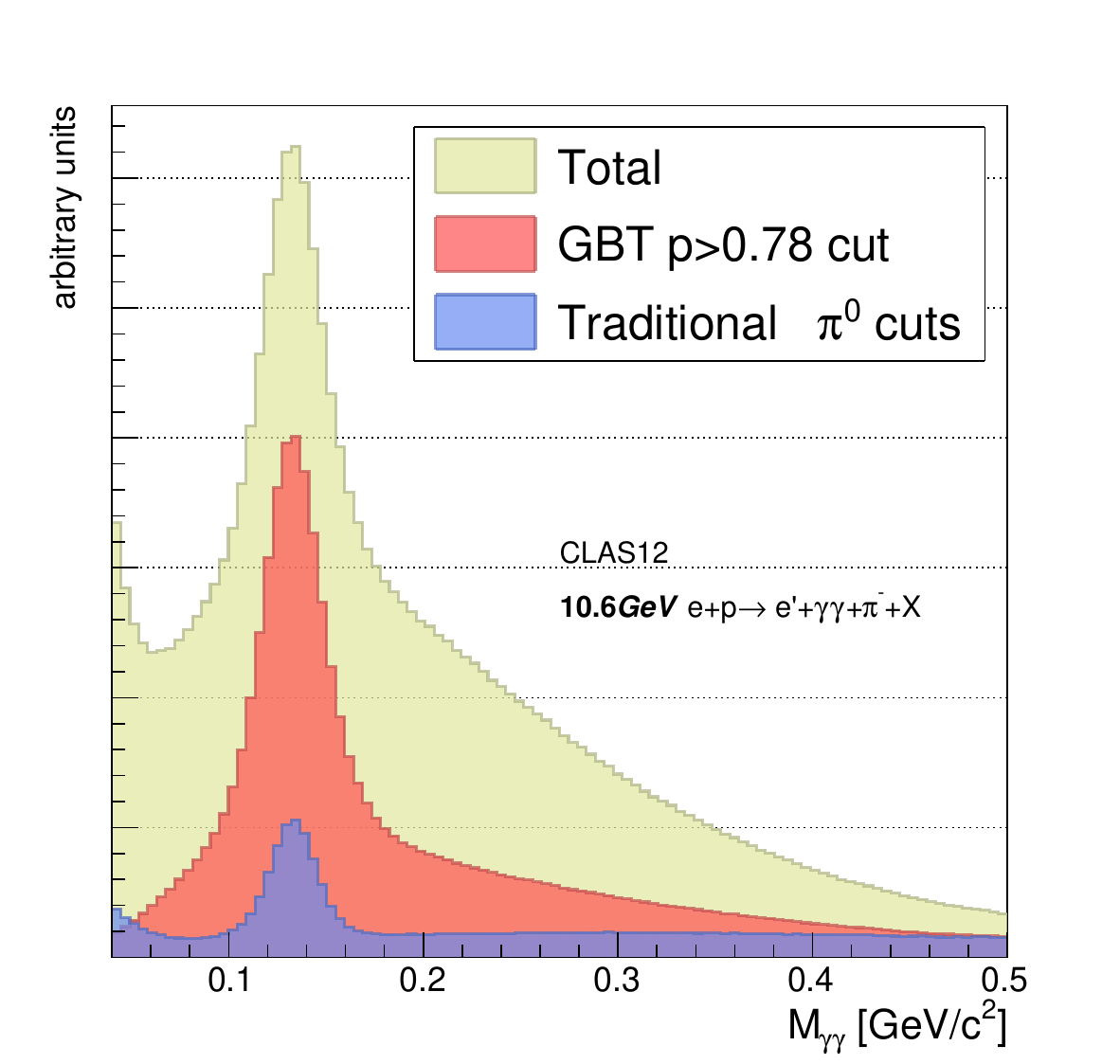}
    \caption{CLAS12 diphoton invariant mass spectrum. The legacy selection (blue) keeps photons with $E_\gamma>0.6$ GeV, whereas the new Gradient Boosted Tree vetoes photons with classifier score $p<0.78$ (red).}
    \label{fig:diphoton_dists}
\end{figure}

SIDIS events are selected by requiring a minimum squared four-momentum transfer $Q^2>1$ GeV$^2/c^2$ and minimum final state hadronic mass $W>2$ GeV$/c$. Radiative effects are minimized with a threshold cut $y<0.8$ on the fractional energy lost by the electron when scattering. A $P_1>1.25$ $\mathrm{GeV}/c$ cut is placed on charged pions to improve four-momentum resolution. Similarly, candidate $\pi^0$s are built from reconstructed photons satisfying the condition $E_{\gamma}>0.2$ GeV. A Gradient Boosted Tree (GBT) model was trained with dedicated Monte Carlo simulations to reduce contamination from false photons~\cite{matousek2024photonclassificationgradientboosted}. CLAS12's ECals are prone to errors in reconstruction where stray calorimeter hits from wide hadronic showers are independently clustered, resulting in an overabundance of false neutral particles. For each photon, the GBT algorithm leverages nearest-neighbor features (e.g. the relative angles to the nearest hadron or photon) and intrinsic properties (e.g., $E_{\gamma}$) to calculate a $p\in[0,1]$ score, from which a $p>0.78$ cut is applied in this analysis to veto false photons. The diphoton distribution shown in Figure~\ref{fig:diphoton_dists} compares the total spectrum to a subset applying the aforementioned GBT-threshold and another applying a stricter $E_{\gamma}>0.6$ GeV seen in legacy CLAS12 analyses such as in Ref.~\cite{Dilks_2022}. 

Contributions to target fragmentation are reduced by requiring $x_F>0$ for each pion. The variable $x_F$, or x-Feynman, takes a positive value if the outgoing hadron moves in the same direction as the incoming electron in the struck quark center-of-mass frame. To eliminate exclusive events, we apply a cut requiring the missing mass to exceed 1.5 $\mathrm{GeV}/c^2$ and the dihadron $z$ to be no more than $0.95$. 

\begin{figure}[h]   \includegraphics[width=0.45\textwidth]{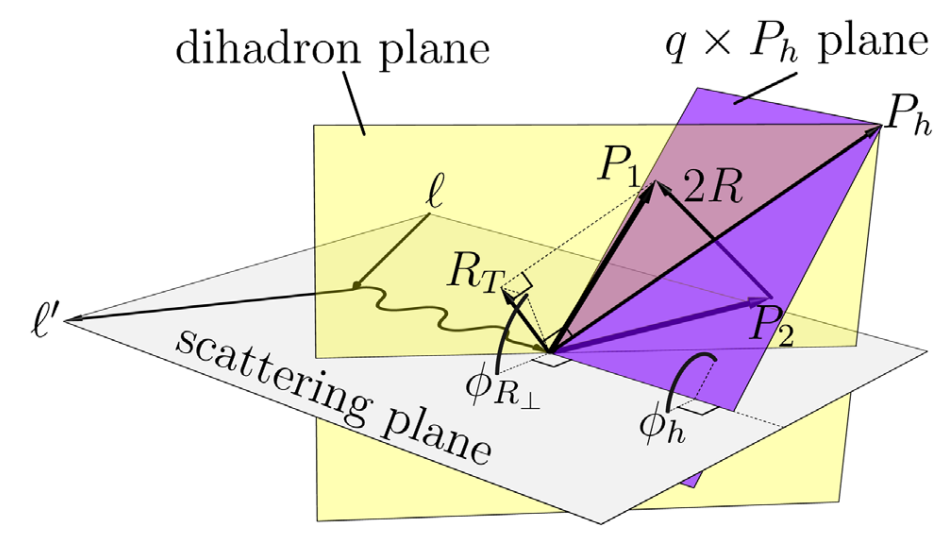}
    \centering
    \caption{The coordinate system for dihadron production. The scattering plane is defined by the incoming and outgoing lepton $\mathbf{\ell}$ and $\mathbf{\ell}'$, the $\mathbf{q}\times\mathbf{P_h}$ plane by the virtual photon $\mathbf{q}$ and total hadron momentum $\mathbf{P_h}$, and the dihadron plane by $\mathbf{q}$ and $\mathbf{R}_T$. The azimuthal angles $\phi_h$ and $\phi_{R_\perp}$ are given by angles spanned between two planes.}
    \label{fig:DiSIDISANGLES}
\end{figure}

The twist-3 PDF $e(x)$ and the DiFF $G_1^\perp$ reveal themselves through sinusoidal modulations of the dihadron azimuthal angles. By measuring beam-spin asymmetries, one isolates those modulations, thereby extracting observables that are directly sensitive to these functions. In Figure~\ref{fig:DiSIDISANGLES}, the angular variables $\phi_h$ and $\phi_{R_\perp}$ are represented 3-dimensionally, with the total hadron momentum $\mathbf{P}_h$ and the relative hadron momentum $2\mathbf{R}=\mathbf{P}_1-\mathbf{P}_2$ with $\mathbf{P}_1$ assigned to the charged pion. They are defined by

\begin{equation}\label{eq:phi_h}
\phi_h=\frac{(\qcl)\cdot\Ph}{|(\qcl)\cdot\Ph|}\,\,\mathrm{arccos}\left(\frac{(\qcl)\cdot(\qcPh)}{|\qcl|\cdot|\qcPh|}\right),
\end{equation}

\begin{equation}\label{eq:phi_RT}
    \phi_{R_\perp}=\frac{(\qcl)\cdot\Rperp}{|(\qcl)\cdot\Rperp|}\,\,\mathrm{arccos}\left(\frac{(\qcl)\cdot(\qcRperp)}{|\qcl|\cdot|\qcRperp|}\right),
\end{equation}

where $\mathbf{R}_\perp$ is given by the projection\footnote{An alternate definition given by $\mathbf{R}_{T}$ projects $\mathbf{R}$ transverse to the dihadron momentum, differing from how $\mathbf{R}_\perp$ projects $\mathbf{R}$ transverse to the fragmenting quark momentum with corrections on the order of $1/Q^2$~\cite{bacchetta_two-hadron_2004}.}

\begin{equation}\label{eq:Rperp}
    \Rperp=\frac{z_{2}\mathbf{P}_{1\perp}-z_{1}\mathbf{P}_{2\perp}}{z_1+z_2},
\end{equation}
$\mathbf{q}$ denotes the momentum of the virtual photon, $\mathbf{P}_h$ the momentum of the dihadron pair, and $\mathbf{l}$ the momentum of the incoming electron.

The beam-spin dependent part of the dihadron cross section can be written as a linear combination of amplitudes multiplied by sinusoidal modulations of $(\phi_h,\phi_{R_\perp})$~\cite{Gliske_Bacchetta_Radici_2014}:
\begin{align}
d\sigma_{LU}&\propto \lambda_e \Bigg[\sum_{m=1}^{2}\Big(A_{LU}^{\sin(m(\phi_h-\phi_{R_\perp}))}\sin(m(\phi_h-\phi_{R_\perp}))\Big)
\nonumber\\+\sum_{m=-2}^{2}&\Big(A_{LU}^{\sin((1-m)\phi_h+m\phi_{R_\perp})}\sin((1-m)\phi_h+m\phi_{R_\perp})\Big)\Bigg].
\end{align}
The amplitudes $A_{LU}^{\sin(m(\phi_h-\phi_{R_\perp}))}$ are at twist-2 and are sensitive to convolutions of the unpolarized PDF $f_1(x)$ and the helicity-DiFF $G_1^{\perp}$ over transverse momentum. In other words,
\begin{equation}
    A_{LU}^{\sin(m(\phi_h-\phi_{R_\perp}))}\propto f_1\otimes G_1^{\perp}.
\end{equation}
At twist-3, the amplitudes $A_{LU}^{\sin((1-m)\phi_h+m\phi_{R_\perp})}$ are sensitive to convolutions of the chiral-odd PDF $e(x)$ and the leading-twist interference fragmentation function $H_1^\sphericalangle$. The $H_1^\sphericalangle$ DiFF correlates the transverse polarization of the outgoing quark to the produced dihadron pair and while it has been extracted using Belle $e^+e^-$ data for $\pi^+\pi^-$ dihadrons~\cite{Vossen:2011PRL,Courtoy:2012PRD}, there is no published extraction of $H_1^\sphericalangle$ for $\pi^\pm\pi^0$ pairs. Since the $\phi_h$ dependence drops for $m=1$, the transverse momentum can be safely integrated out leaving a term still sensitive to $e(x)$. The corresponding amplitude is written as~\cite{bacchetta_two-hadron_2004}
\begin{equation}
A_{LU}^{\sin(\phi_{R_\perp})}\propto\left[x\,e(x)H_1^\sphericalangle(z,M_h)+\frac{1}{z}f_1(x)\tilde{G}^{\sphericalangle}(z,M_h)\right].
\end{equation}
The second term multiplies the well-constrained $f_1(x)$ to the twist-3 DiFF $\tilde{G}^\sphericalangle$ which spectator models show to be small compared to $H_1^\sphericalangle$~\cite{PhysRevD.99.054003}. Ongoing analyses aim to extract $\tilde{G}^\sphericalangle$ point-by-point, which will not only provide the first measurement of a twist-3 fragmentation function but also constrain its impact on $e(x)$ extractions~\cite{Avakian:2020JLab}.


Beam-spin asymmetries for $\pi^{\pm}\pi^{0}$ dihadrons must be corrected for background from random $\gamma\gamma$ pairs that mimic $\pi^{0}\to\gamma\gamma$.  We assume the background asymmetry $A^{\text{bkg}}_{LU,\ell}$ is independent of the diphoton invariant mass.  A side-band region, $M_{\gamma\gamma}\in[0.20,0.40]~\text{GeV}/c^2$, therefore provides a clean sample of pure combinatorial background.  Using a log-likelihood fit, we first extract $A^{\text{bkg}}_{LU,\ell}$ for every modulation $\ell$ in this region.

Inside the signal window $M_{\gamma\gamma}\in[0.106,0.166]~\text{GeV}/c^2$ we extend the event-by-event likelihood to
\begin{equation}
\mathcal{P}_{\pm}(\phi_h,\phi_R)=1\pm P_b\!\sum_\ell\psi_\ell\,\!\bigl[u\,A^{\text{sig}}_{LU,\ell}+(1-u)\,A^{\text{bkg}}_{LU,\ell}\bigr],
\end{equation}
where $\psi_\ell\equiv\psi_\ell(\phi_h,\phi_R)$ represents an independent sinusoidal modulations and the purity $u\equiv u(\phi_h,\phi_R)$ is the probability that a given event truly contains a reconstructed $\pi^{0}$.  The purity is obtained in each kinematic bin by fitting the $M_{\gamma\gamma}$ spectrum with a Gaussian (signal) plus a fourth-order polynomial (background) and then evaluating the signal-fraction in 10$\times$10 sub-bins of $(\phi_h,\phi_R)$. During the final likelihood minimization the previously fixed $A^{\text{bkg}}_{LU,\ell}$ values and the binned purities $u(\phi_h,\phi_R)$ leave only the desired signal asymmetries $A^{\text{sig}}_{LU,\ell}$ as free parameters.  Monte Carlo injection tests verify that this piecewise purity weighting provides the most accurate extraction.


\begin{figure*}[ht]
    \includegraphics[width=0.95\textwidth]{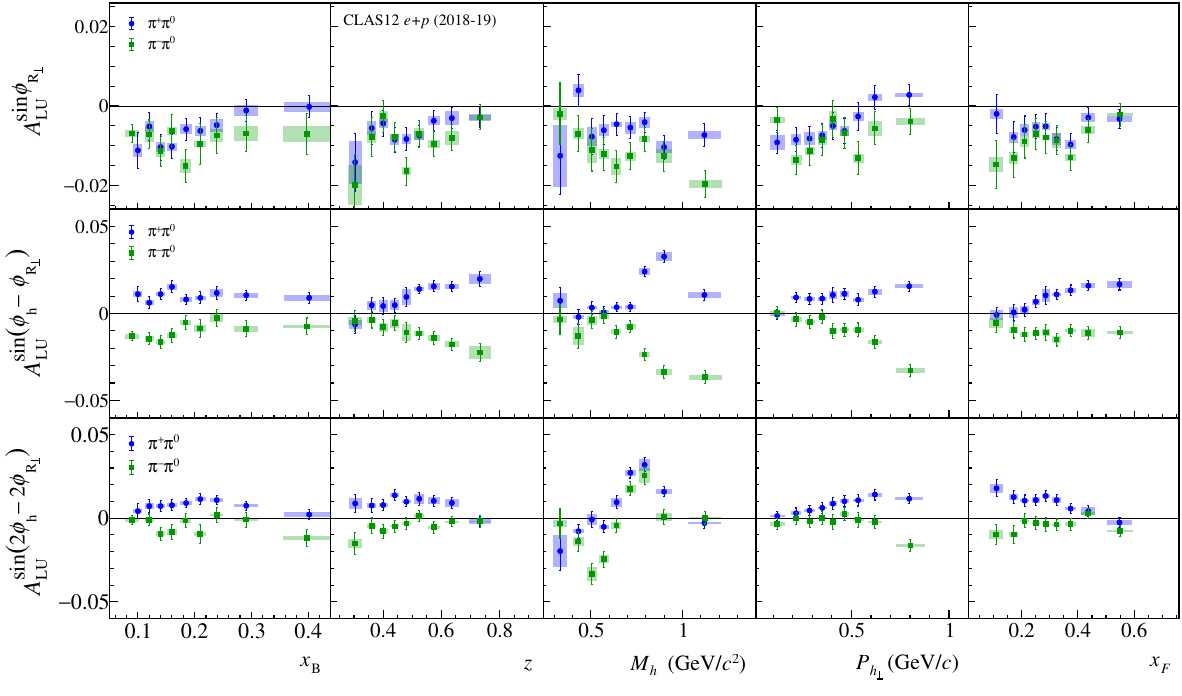}
    \caption{Results for the $A_{LU}^{\smash{\sin\phi_{R_\perp}}}$ (top row), $A_{LU}^{\smash{\sin(\phi_h-\phi_{R_\perp})}}$ (middle row) and $A_{LU}^{\smash{\sin(2\phi_h-2\phi_{R_\perp})}}$ (bottom row) as a function of $x$, $z$, $M_h$, $P_{h\perp}$ and $x_F$ for $\pi^+\pi^0$ (blue circles) and $\pi^-\pi^0$ (green squares) dihadrons. The error bars represent the total uncertainty. The boxes' extents show the systematic uncertainties for each point.}
    \label{fig:asym_grid}
\end{figure*}

The $\pi^+\pi^0$ and $\pi^-\pi^0$ beam spin asymmetries are measured as functions of $x$, $z$, $M_h$, $P_{h\perp}$ (the transverse momentum of the dihadron pair with respect to the scattering plane) and $x_{F}$. Systematic errors in these measurements were investigated using a Monte Carlo simulation that combined the CLAS12-tuned PEPSI generator~\cite{Mankiewicz:1991dp} with a GEANT4-based detector simulation~\cite{agostinelli_geant4simulation_2003, ungaro_clas12_2020}. The leading contributions were found to be those from bin migration, the determination and azimuthal-binning of the $\pi^0$-purity, and the scale error on the experiment's beam polarization. Point-to-point bin migration corrections are largest for the rapidly-changing $M_h$ binned asymmetries where on average $25-35\%$ of simulated dihadrons migrate from one kinematic bin to another after reconstruction. Injection studies showed that strong correlations between the event purity and azimuthal angles led to significant pulls when extracting asymmetries. To mitigate this effect, fine binning schemes of $u(\phi_h,\phi_{R_\perp})$ were adopted, and an additional point-to-point systematic uncertainty was included based on sampling hundreds of different values for the event-by-event purity. A scale uncertainty of $3.0\%$ was set for the beam polarization.

Additional sources of systematic effects were explored, including particle misidentification, baryon decays, the definition of the purity fit's background function and the sideband regions. Overall these contributions were negligible, apart for the latter at relatively low values of $M_h$. Asymmetries extracted from data selected using the photon GBT classifier cuts were generally consistent with those from the legacy analysis cuts within statistical uncertainties, except for modulations sensitive to $\rho$ resonances, where the GBT-based method yields a purer sample of true $\pi^0$s.

Figure~\ref{fig:asym_grid} shows $A_{LU}^{\smash{\sin\phi_{R_\perp}}}$ (top row), $A_{LU}^{\smash{\sin(\phi_h-\phi_{R_\perp})}}$ (middle row) and $A_{LU}^{\smash{\sin(2\phi_h-2\phi_{R_\perp})}}$ (bottom row) as a function of $x$, $z$, $M_h$, $P_{h\perp}$ and $x_F$ for $\pi^+\pi^0$ and $\pi^-\pi^0$ dihadrons. The $\pi^+\pi^0$ results are consistent with earlier CLAS12 measurements~\cite{Dilks_2022}, while the use of the photon GBT classifier in the present analysis provides a five-fold increase in statistics.

The $A_{LU}^{\smash{\sin\phi_{R_\perp}}}$ for both dipion channels integrated across all kinematics reveals a statistically significant ($>5\sigma$) overall negative asymmetry. We find that the $\pi^+\pi^0$ channel's asymmetry becomes consistent with zero at $x>0.3$. As this asymmetry is proportional to $e(x)H_1^\sphericalangle(z,M_h)$, both signals confirm a nonzero PDF $e(x)$, complementing the results from $\pi^+\pi^-$ dihadrons~\cite{PhysRevLett.126.152501}. The $\pi^\pm\pi^0$ channel's $\approx -1\%$ asymmetry contrasts with the larger $\approx +4\%$  asymmetry observed in the $\pi^+\pi^-$ channel, indicating a sign- and magnitude-dependence for $H_1^\sphericalangle(z,M_h)$ on the outgoing pion pair. While point-by-point extractions of $e(x)$ using this data are contingent on future measurements of $H_1^\sphericalangle(z,M_h)$ for $\pi^\pm\pi^0$ pairs, recent quark-jet model predictions provide insight to the DiFF's relative sign for different dipion final states~\cite{Matevosyan_Kotzinian_Thomas_2018}.

Vector meson decays (ex: $\rho^\pm\rightarrow\pi^\pm\pi^0$) play an influential role in shaping both single and dihadron asymmetries~\cite{Avakian2025CLAS12SSA}. Signatures of $\rho$ decay are visible in both channels' $A_{LU}^{\smash{\sin(\phi_h-\phi_{R_\perp})}}$ measurement as a function of $M_h$. The amplitude is sensitive to the $\ket{\ell,m}=\ket{1,1}$ DiFF partial wave\footnote{Subscripts $OT$ indicate an interference of an unpolarized dihadron and transversely polarized dihadron at the amplitude level~\cite{bacchetta_two-hadron_2004}.} $G_{1,OT}^\perp$ and is seen to exhibit a positive enhancement for $M\geq M_{\rho^\pm}$. Spectator model calculations predict this enhancement in $\pi^+\pi^-$ pairs due to an interference of $s$- and $p$-wave dihadrons~\cite{Luo_Sun_2020}. A clear sign-difference in the $\pi^+\pi^0$ and $\pi^-\pi^0$ signal is observed, marking the first experimental evidence of isospin-dependence of $G_1^\perp$. 

Another striking $\rho$ enhancement is measured in $A_{LU}^{\smash{\sin(2\phi_h-2\phi_{R_\perp})}}$ as a function of $M_h$. Spectator model calculations for $\pi^+\pi^-$ in the $\ket{\ell,m} = \ket{2,2}$ DiFF partial wave $G_{1,TT}^{\perp}$ predict this asymmetry to peak near $M_h=M_\rho$ due to the interference between two \mbox{$p$-wave} dihadrons sharing the same transverse polarization~\cite{Luo_Sun_Xie_2020}. Our results for $\pi^\pm\pi^0$ dihadrons strongly resemble this effect, with both channels exhibiting a large $\approx +3\%$  asymmetry at $M_h\approx0.75$ $\text{GeV}/c^2$. The positive signal also mirrors the CLAS12 results for $\pi^+\pi^-$ dihadrons~\cite{timothy_thesis} and is a direct contrast to the sign-difference observed in our $\ket{\ell,m}=\ket{1,1}$ asymmetries.  

In conclusion, this Letter presents the first beam spin asymmetry measurements of $\pi^+\pi^0$ and $\pi^-\pi^0$ SIDIS dihadrons. Statistics for both channels were quintupled compared to legacy CLAS12 $\pi^0$ analyses by implementing a high-purity Gradient Boosted Tree photon classifier. Small negative ($\approx -1\%$) $\sin\phi_{R_\perp}$ asymmetries, sensitive to the twist-3 PDF $e(x)$, were measured for both channels. These measurements provide further evidence that twist-3 dynamics are not a negligible correction but an essential part of the proton's internal structure. These asymmetries will help to isolate the elusive distribution, delivering new insight into quark–gluon correlations and offers an additional avenue for its point-by-point extraction. Enhancements of the $\sin(\phi_h-\phi_{R_\perp})$ and $\sin(2\phi_h-2\phi_{R_\perp})$ BSAs near the $\rho^\pm$ mass were observed, indicating the presence of resonance-like structures in the helicity DiFF $G_1^\perp$. Future studies will perform a full partial wave decomposition of the $\pi^\pm\pi^0$ beam spin asymmetries and extend these methods to the remaining dipion channels. 

\section*{Acknowledgments}
This material is based upon work supported by the U.S. Department of Energy, Office of Science, Office of Nuclear Physics under contracts DE-SC0024505 and DE-AC05-06OR23177. This research was supported by the National Science Foundation (NSF) under grant agreement DGE-2139754. We extend our gratitude to the DOE and NSF for their financial support, which made this work possible. We thank and acknowledge the support of the Deutsche Forschungsgemeinschaft (DFG).

\end{document}